# Recommendations for Government Development and Use of Advanced Automated Systems to Make Decisions about Individuals

March 1, 2024

Susan Landau, James X. Dempsey, Ece Kamar, Steven M. Bellovin

Contestability—the ability to effectively challenge a decision—is critical to the implementation of fairness. In the context of governmental decision making about individuals, contestability is often constitutionally required as an element of due process; specific procedures may be required by state or federal law relevant to a particular program. In addition, contestability can be a valuable way to discover systemic errors, contributing to ongoing assessments and system improvement.

The ability of a person to challenge a decision based in whole or part on an automated system can be limited due to insufficient information about the system, the technical opaqueness of the system, or broader problems with the right to contest as it exists on the books and on the ground (e.g., Citron, Wexler). As automated systems become more advanced due to the incorporation of machine learning and other artificial intelligence technologies, contestability may become even more difficult to achieve.

Yet it is not inevitable that advances in automated decision making will necessarily make contestability more difficult. To the contrary, as researchers and developers have shown, conscious choices made in system design can ensure that advanced automated systems enable meaningful contestability, perhaps even better than current systems do. Here, we provide recommendations for the development and use of advanced automated systems to enable contestability.[1]

Contestability is not merely a best practice. Where decisions are being made about individuals, contestability is a requirement. Building on Executive Order 14110 on Safe, Secure and Trustworthy Development and Use of Artificial Intelligence (Oct. 30, 2023), the federal government should adopt binding standards on contestability applicable to the design, testing, implementation, and monitoring of advanced automated systems used by the federal government or in the administration of federally funded programs.

Note that we use "system" here to encompass both technology and humans and both policy and data. Thus, when we say "system," we are encompassing such disparate systems as a hiring system, a system for allocating police resources, a system for granting or denying veterans' benefits or determining Medicaid eligibility, or a system for prioritizing enforcement actions. Note also that we use "program" here to refer to a governmental benefit, service, function or activity (e.g., a veterans' benefits program or an enforcement program), not to computer code or software.

In some systems, the automated technology will be public-facing and its impact on individuals will be immediately apparent, and the need for contestability will be clear. In other contexts, however, the automated technology will be operating in the background, supporting decisions that directly affect individuals. In such cases, careful consideration will need to be given to transparency to ensure that individuals can discern whether and how technology is affecting them, as a premise for contesting the results of such decision-support functions. In yet other contexts, automated analytics capabilities will improve backend operations of government agencies without implicating individual rights.

---

[1] When it comes to contestability, not all AI/ML techniques or capabilities are equal. Some may be incompatible with contestability; depending on the context, these should not be used.

On January 24-25, 2024, with support from the National Science Foundation and the William and Flora Hewlett Foundation, we convened a diverse group of government officials, representatives of leading technology companies, technology and policy experts from academia and the non-profit sector, advocates, and stakeholders for a workshop on advanced automated decision making, contestability, and the law. Informed by the workshop's rich and wide-ranging discussion, we offer these recommendations. A full report summarizing the discussion is in preparation.

**Contestability by Design**

Contestability is often discussed in terms of transparency, interpretability or explainability of the automated decision system, and contestability does require that meaningful explanations be provided to the user and the subject. However, contestability is more than just transparency, interpretability, or explanation. Moreover, the definitions of those terms are unsettled. In particular, "explanation" may have different meanings in AI/ML science versus in law or social science (see, e.g., Mittelstadt et al.).

To ensure contestability, those involved in the development of a system need to understand what can go wrong (what are the types and sources of errors) in a decision-making process.[2] Sometimes, a decision is wrong because the system was trained on inaccurate, biased, or irrelevant data. Other times, the automated elements of a system do not properly reflect the law, policy, or rules applicable to the particular government or private-sector program or activity. Sometimes the system is wrong in a specific decision because the data it uses about an individual is inaccurate or incomplete or accurate but irrelevant. Sometimes the system is wrong because a machine recommendation is unclear, and the human misinterprets the recommendation, making a wrong decision. Another possibility is that the tool is probabilistic and produces a certain percentage of errors even when everything is done perfectly.[3] Designing for the individual right to contestability should take into account all of these (and possibly other) error types. Nonetheless, although designing advanced automated systems to support a meaningful right of contestability is difficult, it is not impossible—and *it is often required by law.*

---

[2] In referring to a system outputs as being "wrong" or "erroneous," we have in mind the language of the federal Administrative Procedure Act: "arbitrary, capricious, an abuse of discretion, or otherwise not in accordance with law; … in excess of statutory … authority, or limitations, or short of statutory right; … unsupported by substantial evidence …; or unwarranted by the facts … ." 5 U.S.C. § 706.

[3] Designers and users alike need to be aware of the error rate in such systems.

Contestability must be incorporated into a system by design, and considerations related to contestability must be addressed when initially determining whether to include automated decision making capabilities in a system and then throughout the development lifecycle, from conceptualization to implementation and must be examined after deployment in ongoing monitoring and improvement. At the same time, it is important to understand what level of contestability is sufficient for a given domain, and design accordingly. Note that no single practice or feature can ensure adequate contestability. Instead, contestability can be achieved only though application of a series of techniques, starting with design and impact assessment. Some of the techniques for ensuring contestability will relate to the technology, while others will relate to overall system policy.

***Recommendations: Building Contestability into Advanced Automated Systems Used in Government Decision Making about Individuals***

**Recommendation 1: In contexts where contestability is required, government should ensure that adequate notice is provided that an automated decision-making system is being developed and being used.** Notice is an essential prerequisite of contestability; this means both (i) notice to the public before the decision is made to adopt automated decision making for a system and then consultation as it is being developed and (ii) notice to individual subjects that their case has been decided based in whole or part on an automated process.

**Notice to the public must be adequate to allow for systemic challenges to a system**, which can identify problems before large numbers of individuals are unfairly affected.

**Notice to individuals must be understandable;** thus, notice must provide sufficient detail so that affected individuals and their representatives or advocates can understand how a decision was made and what a person must present to contest it.[4]

The degree of transparency necessary to support contestability will vary by context. Especially where the government fails to justify an outcome in a way that is understandable by the affected individual and his or her representative, it may be necessary for advocates and litigators to delve into how the system was constructed. Datasheets or model cards as documentation for how a system was built could enable contestability (Mitchell et al., Ehsan et al.), but in some cases a deeper examination of methods, criteria, code and data may be necessary along with expert analysis by those asserting challenges.

---

[4] In *Goldberg v. Kelly,* 397 U.S. 254, 267 (1970), the Supreme Court specified that "[t]he opportunity to be heard must be tailored to the capacities and circumstances of those who are to be heard." Thus the written notice that undergirds due process must be "in terms comprehensible to the claimant." *Ortiz v. Eichler*, 616 F. Supp. 1046, 1061 (D. Del. 1985), *aff'd*, 794 F.2d 889 (3d Cir. 1986). "An elementary and fundamental requirement of due process…is notice reasonably calculated, under all the circumstances, to apprise interested parties of the pendency of the action and afford them an opportunity to present their objections." *Mullane v. Cent. Hanover Bank & Tr. Co.*, 339 U.S. 306, 314 (1950).

**Recommendation 2: Contestability must be incorporated into the system design, beginning with the decision whether to use an advanced automated system in a decision-making or a decision-supporting role.** A risk assessment provides one opportunity early in the development lifecycle to surface issues of contestability.

Design choices—including decisions about which type of advanced automated system to deploy—can affect the ability of individuals or their representatives to effectively contest a decision based in whole or part on automated processes. If individuals can face adverse consequences (such as denial of a federal loan, loss or reduction of benefits, or increased scrutiny in a policing or enforcement context) from the decisions made or recommended by a system, then system designers must choose a technique or technology understandable (e.g., interpretable) by affected individuals and their representatives (if any).

**Recommendation 3: Designers should always consider the option of *not* deploying an advanced automated decision-making technique or technology, and they must build into each system affecting individuals the option of an off-ramp** (that is, the ability, even though it is a very hard decision to make if considerable resources have been expended in development, to exit from the use of automated decision making) if a system is determined, after deployment, to be not sufficiently contestable. Similarly, agencies should carefully monitor the volume and nature of challenges to decisions made or aided by deployed systems; a high rate of challenges may suggest flaws in the system, requiring redesign or even decommissioning.

**Recommendation 4: Design consultations and testing should include different types of system participants, including operators, end users, decision makers, and decision subjects.**[5] Managers familiar with the legal and policy aspects of a governmental program or function must work together with experts in AI or other relevant advanced analytic technologies at all stages of system lifecycle from initial conceptualization through post-deployment assessment and revision.

To ensure contestability, there must be participation throughout a project's life cycle not only from experts in machine learning but also data scientists, statisticians, and experts in such fields as human-computer interface (HCI), design, sociology, cognitive psychology, linguistics, law, and public policy. Other areas of expertise, including criminology, health care, and economics, may also be needed depending on the type of system being developed.

**Recommendation 5: Stakeholders who will be directly affected by an advanced automated system must be involved or represented at all stages of its development and use.** This includes from the initial discussions of the scope of the system, through design, development, iterative testing phases, deployment, post-deployment assessment, and revisions/updates of the system.[6] Special effort must be

---

[5] See NIST RMF Playbook, Measure 2.8.
[6] In the case of systems affecting children and perhaps some other categories of persons, participation can be through representatives or advocates.

made to include persons who are disabled, lack English language skills, or have otherwise traditionally been disadvantaged from full participation in government processes.[7] Because they could offer unique systemic perspectives, advocates such as lawyers handling challenges in the program or function at issue should also be included.

**Recommendation 6: The contestability features of a system must be stress tested with real world examples and scenarios before field deployment.**[8] This also holds true for non-trivial modifications of a deployed system. Before field deployment, it is critical to conduct pilots on real data and in consultation with individuals in all the different types of actual communities in which the system will be deployed. Among other issues, testing should consider whether there are misunderstandings between technology developers on the one hand and lay users and data subjects on the other hand. After deployment, systems may be updated and the scenarios in which they are used may change, so there must be ongoing evaluation of the system's operation.

**Recommendation 7: Contestability processes should be equally accessible and usable by people with different backgrounds,** including—but not limited to—different cultures, languages, education levels, and incomes. Having robust contestability processes is not sufficient in itself; those processes must be widely accessible and usable.

Contestability of any system depends in large part on who will be contesting decisions, what resources they have available to them, and what barriers they face. Governments already struggle to provide clear explanations of what people need to prove to get or keep benefits. Low-income people especially find it difficult to contest government decisions such as benefit cuts or denials. Many affected persons do not have the money needed to hire a lawyer, and the capacity of organizations providing free legal services is severely limited. Even without adoption of advanced technology, formal processes to contest can be confusing, involve burdensome paperwork requirements, and require time away from work or caregiving obligations. As automated systems are designed and implemented, government entities should simplify, streamline and explain more clearly their application and appeal processes.

**Recommendation 8: Reproducibility is crucial.** In order to be contestable, a system must be able to reproduce a given decision (that is, the same inputs must produce the same outputs at the time of challenge). Thus, there needs to be version control and thorough recordkeeping of the systems being used (Aler Tubella et al.), including of the parameters of the models created from training data.

Governments should fully document the relevant data (including training data), statistical modelling, impact projections, system changes, and assumptions involved in the design, development, and testing of any system prior to deployment. This information will aid the agency in making decisions about how the system will be deployed, if at all. In addition, it will aid post-deployment accountability efforts by allowing for internal and public review, strengthening appeal processes, and facilitating agency or legislative oversight efforts.

---

[7] Though there is expertise is available in how to conduct inclusive project design consultations, it is critical that such efforts be done in a way that actually engages individuals who traditionally were not involved in such consultations (see, e.g., Sloane, 2022).

[8] Individuals in the test set should not end up worse off than if their case had been handled by the old system.

**Recommendation 9: The automated features of a system should never be allowed to supplant or displace the criteria specified in law for any given program or function.** The convenience of the programmers or even the technical possibilities of automated systems must take a back seat to what the law requires. For example, if the legal standard is "medical necessity," the factors or criteria considered by the automated process should not be presumed to be the only way to demonstrate medical necessity. If an individual has a due process right to contest a decision, that must include the right to present to the reviewing authority factors or criteria relevant to the legal standard that were not included in the automated process.

**Recommendation 10: Additional research could help government agencies design and implement meaningful contestability processes.** Such work might include a clearer understanding of understandability, better knowledge of the social impact of advanced automated decision making on various communities (including businesses and organizations), whether new legal theory is needed to ensure contestability of automated systems, and evaluation of the risks of "out-of-control" contestability (e.g., when AI bots create a deluge of comments to government agencies in response to a rulemaking or other public consultation).

Although new research could improve the design and implementation of contestability in advanced automated decision-making systems, this does not obviate any of our other recommendations; the actions we recommend must proceed even as contestability capabilities improve due to new understanding and research.

**Contestability Must be Addressed in the Procurement Process**

The procurement process—the nuts and bolts of government contracting—is critical because many automated decision-making systems will be designed and built (and may be managed as a service) for the government by contractors. Thus, the procurement process must be part of the government's efforts in ensuring contestability.

**Recommendation 11: The procurement process should be leveraged to ensure that advanced automated systems genuinely enable contestability**. Solicitations and contracts must clearly require contractors to deliver contestability as a core system feature. Contractors should not be allowed to use assertions of trade secrecy or other intellectual property claims to frustrate contestation.[9] As OMB develops procurement guidelines for AI pursuant to Section 10.1(d)(ii) of the AI Executive Order, it should

---

[9] The logic of an advanced automated process may or may not be useful for contestability, but that question should not be pretermitted by IP claims.

include contestability as a core element in procurement processes; a set of unified guidelines that all procurement officers must adhere to across agencies would be powerful and actionable.

**Recommendation 12: Federal officials should ensure that contestability is required of the states implementing federal programs and of private companies whose systems, such as credit scoring, are used by the government in contexts affecting individuals.** Federal officials can do this by using their approval and oversight authority over programs that are administered partly by states or receive federal funds.

**Recommendation 13: The government should develop expertise to ensure rights-impacting advanced automated systems are contestable.**

The type of socio-technical expertise needed by agencies designing, procuring, and using advanced automated systems is hard to come by. Training within individual agencies, as called for under EO 14110, may be unnecessarily narrow—and thus ultimately ineffective—given the rapid ongoing evolution of the technology, the constitutional and thus cross-disciplinary foundations of contestability, and the generalizable nature of processes for stakeholder consultation as well as for government procurement. It is necessary to ensure that the government workforce has sufficient understanding regarding the development and use of these socio-technical systems.

**Recommendation 14: The federal government should develop a centralized training function in development and assessment of advanced automated system for government programs.**[10] Among others in government who will need to understand the implications of advanced automated systems, agency staff who will be adjudicating appeals, including administrative law judges, should be specially trained to identify the risks of automated systems.

**Recommendation 15: There should be formal—and informal—ways set up to ensure sharing knowledge gained in the development, procurement, and use of these systems.** Many of the efforts within federal, state, and local governments will have common characteristics in terms of the communities they are serving and the challenges they face. The Chief AI Officers of the agencies who will be appointed pursuant to Section 10 of EO 14110 should take this on as one of their responsibilities,

---

[10] This recommendation goes beyond the provisions on AI talent in EO 14110 on Safe, Secure and Trustworthy Development and Use of Artificial Intelligence (Oct. 30, 2023). Sections 10.2 (a)-(f) of the EO address measures the government will take in recruitment and hiring, but the talent pool will always be too small, as universities and other educational institutions are unlikely by themselves to produce enough graduates with the skills needed by the private sector and government. Likewise, Section 5.1 of the EO seek to improve the nation's ability to attract AI talent from abroad, but there too the U.S. is in competition with government of other nations and the private sector of other countries also facing the same talent shortage. Section 10.2(g) of the EO recognizes that the government will have to undertake training itself, but it leaves that training to each agency head. Just as the federal government has established centralized training facilities for other skills, such as languages, cryptography, and law enforcement, it should establish a centralized AI governance institute.

and the interagency council established under the same section of the EO should serve as one vehicle for such knowledge sharing.

**Recommendation 16: At all levels of an agency, from the agency head to the procurement officer to the case worker or other person interfacing with affected individuals, officials need to understand, at a level appropriate to their roles, the benefits, limitations, and risks of automated decision making.** Likewise, at all levels, officials need to understand, at a level appropriate to their roles, the capabilities of automated decision making to deliver on stated goals. That presumes that officials are clear on their goals for the system being developed (for example, whether the intent is to enroll more people in a program or fewer).

**Recommendation 17: In order to build sufficient expertise within agencies, as an early effort, agencies should attempt some low-risk, high-gain systems.** A low-risk system is one that does not have adverse effects on individuals if there are malfunctions.

**Resources**

Aler Tubella, A., Theodorou, A., Dignum, V., et al. (2020). Contestable black boxes. In V. Gutiérrez-Basulto, T. Kliegr, A. Soylu, et al. (Eds.), *Rules and reasoning* (Vol. 12173). Springer.

Citron, Danielle Keats, Technological Due Process, 85 *Wash. U. L. Rev*. 1249 (2008).

Ehsan, E., Liao, Q.V., Passi, S., Riedl, M.O., and Daume, H., Seamful XAI: Operationalizing Seamful Design in Explainable AI, 2022.

Engstrom, David Freeman, Ho, Daniel E., Cuéllar, Mariano-Florentino, Sharkey, Catherine M., Government by Algorithm: Artificial Intelligence in Federal Administrative Agencies, 2020.

Lyons, Henrietta, Velloso, Eduardo, and Miller, Tim, Conceptualising Contestability: Perspectives on Contesting Algorithmic Decisions, roc. *ACM Hum.-Comput. Interact*., Vol. 5, No. CSCW1, 2021.

Mitchell, M., Wu, S., Zaldivar, A., Barnes, P., Vasserman, L., Hutchinson, B., Spitzer, E., Raji, I.D., and Gebru, T., Model Cards for Model Reporting, *Proceedings of the Conference on Fairness, Accountability, and Transparency, 2019*;

Mittelstadt, Brent, Russell, Chris, and Wachter, Sandra, Explaining Explanations in AI, *FAT 19,* 2019.